\documentclass{amsart}

\usepackage{amsmath}
\usepackage{amsthm}
\usepackage{amsfonts}
\usepackage{amssymb}
\usepackage{graphicx}
\usepackage{xcolor}
\usepackage{setspace}
\usepackage{subcaption}
\usepackage{etoolbox}
\usepackage[foot]{amsaddr}
\usepackage{etoolbox}
\usepackage{tikz}
\usepackage{tcolorbox}

\usepackage[printwatermark]{xwatermark}
\newsavebox\mybox
\savebox\mybox{\tikz[color=red,opacity=0.2]\node{DRAFT};}
\usepackage[authoryear,compress]{natbib}

\usepackage{hyperref}
\hypersetup{
    colorlinks,
    linkcolor={red!50!black},
    citecolor={blue!50!black},
    urlcolor={blue!80!black}
}

\makeatletter
    \patchcmd{\NAT@test}{\else \NAT@nm}{\else \NAT@nmfmt{\NAT@nm}}{}{}
    \DeclareRobustCommand\citepos
        {\begingroup
        \let\NAT@nmfmt\NAT@posfmt% ...except with a different name format
        \NAT@swafalse\let\NAT@ctype\z@\NAT@partrue
        \@ifstar{\NAT@fulltrue\NAT@citetp}{\NAT@fullfalse\NAT@citetp}}
    \let\NAT@orig@nmfmt\NAT@nmfmt
    \def\NAT@posfmt#1{\NAT@orig@nmfmt{#1's}}
\makeatother

\title[Est-ce que Vous Compute?]{Est-ce que Vous Compute? \\ Code-Switching, Cultural Identity, and AI}

\author{Arianna Falbo$^{1}$ \\ Travis LaCroix$^{2}$}

\address{\tiny{$^1$Department of Philosophy \\ Brown University}}
\address{\tiny{$^2$Department of Philosophy \\ Dalhousie University}}

\email{arianna\_falbo@brown.edu \textrm {(Falbo);} tlacroix@dal.ca \textrm{(LaCroix)}}

\date{Draft of \today}

\begin{document}

%\begin{center}
%    {{\it  \textbf{Feminism, Social Justice, and AI Special Issue Submission.} \vspace{1cm}}}
%\end{center}

\maketitle
\pagestyle{empty}
\thispagestyle{empty}

\begin{abstract}
\singlespacing
    
    \phantom{a} Cultural code-switching concerns how we adjust our overall behaviours, manners of speaking, and appearance in response to a perceived change in our social environment. We defend the need to investigate cultural code-switching capacities in artificial intelligence systems. We explore a series of ethical and epistemic issues that arise when bringing cultural code-switching to bear on artificial intelligence. Building upon \citepos{Dotson2014} analysis of testimonial smothering, we discuss how emerging technologies in AI can give rise to epistemic oppression, and specifically, a form of self-silencing that we call \textit{cultural smothering}. By leaving the socio-dynamic features of cultural  code-switching  unaddressed, AI systems risk negatively impacting already-marginalised social groups by widening opportunity gaps and further entrenching social inequalities.
    
    \phantom{a}

    \noindent \textbf{\textit{Keywords}} --- cultural code-switching, artificial intelligence (AI), natural language processing (NLP), silencing, epistemic oppression, AI ethics, social dynamics  
\end{abstract}

\section{Introduction}
    \label{sec:Introduction}

    In linguistics, \textit{code-switching} describes using two (or more) languages (or language varieties) within a single text, conversation, or utterance.%
            \footnote{The term `code-switching' (sometimes referred to as `codemixing', `codeshifting', `language alternation', `language mixture', or `language switching') appears in the 1950s; however, observations of these linguistic phenomena in academic writing predate this baptism by several decades; see \citet{Benson-2001} for an historical discussion. Code-switching, like many other communicative phenomena, appears to be governed by several linguistic and extra-linguistic features; see discussion in, e.g., \citet{Gumperz-1977, Pfaff-1979, Poplack-1980, Benson-2001}.}
    This purely linguistic sense of code-switching has been widely discussed in the context of artificial intelligence (AI)---especially for natural language processing (NLP) models.%
            \footnote{See \citet{Jose-et-al-2020} for a recent survey.} 
    AI/NLP research on code-switching typically concentrates on the morphosyntactic features of code-switched linguistic data at the sentential (or sub-sentential) level. However, there is a broader characterisation of code-switching phenomena that is primarily {\it social}, namely, {\it cultural} code-switching \citep{Falbo-2021}. Cultural code-switching relates to how we adapt our overall behaviour, manner of speaking, and appearance in response to a perceived change in our social environment. Unlike the merely linguistic sense of code-switching, cultural code-switching involves `a much more profound shift than toggling between languages or dialects does' \citep[76]{Morton-2019}.\footnote{See also the definitions and related discussions found in \citet{Morton-2014, McCluney-et-al-1979}.} Cultural code-switching is more intimately connected to the self. It is also a mechanism by which we can conform to dominant cultural norms, and, by the same token, it is a means by which we can defect and resist the status quo. 

    As AI-based language technologies continue to improve and become more pervasive in society, it will be increasingly essential that they fluidly interact with humans---i.e., in the way that humans typically interact with one another. However, a significant part of such interactions involves non-linguistic commun\-ication---i.e., the additional features of communication that are relevant for cultural code-switching. Engagement in cultural code-switching is especially common among minority communities, and it may be done for a variety of reasons---conforming to majority-group norms, gaining social acceptance, avoiding hostility or discrimination, or achieving social mobility, to name a few. Accordingly, the extant biases of emerging technologies, which typically arise from dominant-group conventions, may be further exacerbated.\footnote{It is, by now, well-documented that many instances of machine bias result from training, where the technological artefact mirrors extant biases in datasets. See discussion in, e.g., \citet{Friedman-Nissenbaum-1996, Angwin-et-al-2016, O'Neil-2016, Caliskan-et-al-2017, Cave-Dihal-2020, Johnson-2021}.} To mitigate, and not further contribute to, these biases, it is important for researchers in AI---especially those working in the machine learning paradigm and other emerging methods---to take cultural code-switching into account in models of NLP. This is particularly true when these systems are deployed, especially in high-stakes social environments (e.g., legal settings, job interviews, educational institutions) or when they are otherwise integrated into standard consumer-facing technologies or applications.

    The goals of this paper are two-fold, in light of these considerations. First, we motivate the importance of cultural code-switching in research on the ethics of AI. Despite its ubiquity and, more importantly, the unique value and costs that cultural code-switching has for those from marginalised backgrounds, surprisingly little attention has been given to these phenomena, either in the technical literature on NLP models or in the philosophical literature on AI more broadly. This paper bridges this gap by defending the need to investigate cultural code-switching capacities in AI systems (AIS).

    Second, drawing upon the growing literature on epistemic oppression and related work on structural oppression, we canvas the potential moral and epistemic risks involved in implementing---or, more importantly, \textit{failing} to implement---code-switching capacities in AIS. By leaving the socio-dynamic features of cultural code-switching unaddressed, AIS risk negatively impacting already marginalised social groups by widening opportunity gaps and further entrenching social inequalities.%
        \footnote{See \citet{LaCroix-OConnor-2021, Bruner-2017, Bruner-Oconnor-2017, Oconnor-Bruner-2019} for discussion on how the constitution of groups and group dynamics alone may give rise to inequitable conventions.} 
    More generally, reflecting upon the need for cultural code-switching in AIS allows us to enrich our understanding of these phenomena beyond the context of human-AI interaction. Cultural code-switching influences not only first-personal identity---particularly what \cite{Dembroff-Saint-Croix-2019} have called \textit{agential identities}%, which concern how we make our identities publicly available to others
    ---but also broader systems of social coordination within and between groups.

    The paper proceeds as follows. Section~\ref{sec:Background} provides a brief overview of present-day artificial intelligence and machine learning methods, highlighting extant biases that have surfaced. This section also clarifies that the AI research area that is most relevant to code-switching is {\it natural language processing}. However, as mentioned, the code-switching phenomena we are primarily concerned with is broader than the linguistic context alone. Accordingly, section~\ref{sec:Cultural-Code-Switching} makes clear what {\it cultural} code-switching is and some of the reasons we do it, using specific examples.  Section~\ref{sec:Cultural-Smothering} draws upon the literature on epistemic oppression and, in particular, on \citepos{Dotson2014} analysis of \textit{testimonial smothering}. In so doing, we broaden Dotson's framework to include a species of self-silencing, which, following \citet{Falbo-2021}, we call \textit{cultural smothering}. This occurs when cultural code-switching behaviours manifest as a form of self-censoring: one alters aspects of their cultural identity in response to an unwelcoming or hostile social atmosphere. We also discuss how pressures to code-switch in this sense, and thus to engage in or succumb to cultural smothering, often have the structure of a \textit{double bind} choice situation \citep{Frye-1983}. 
    
    Section~\ref{sec:Conclusion} draws some tentative conclusions concerning the relationship between cultural code-switching and emerging technologies. We must be sensitive to how practices of silencing are sustained and reinforced in society and how this maintains unjust social arrangements, thereby limiting access to social goods needed to improve the material conditions of marginalised communities (e.g., safety, education, job opportunities). A natural extension of this carries over to the domain of emerging technologies, particularly AI and NLP. We should be critical of how these same patterns of silencing and epistemic oppression are reproduced and potentially rendered more potent and efficacious in emerging technologies---especially as these technologies become increasingly widespread and seamlessly integrated into everyday life. 
    
     To be clear: it is not our intention to defend any concrete solutions to the problems we identify. We don't offer any optimistic plans for developing or training AI to better avoid or ameliorate epistemic oppression. We are not confident that any such solution ultimately exists. Instead, we offer these arguments as a serious caution, as a lesson concerning the limitations and harms of AI technologies when they are adopted and implemented against the backdrop of a non-ideal, hierarchically structured social world, filled with unjust divisions on the bases of race, gender, sexual orientation, ability, religion, and class, among others. Unless and until the social world changes for the better, we should expect these problems to persist.

\section{Artificial Intelligence, Machine Learning, and Natural Language Processing}
\label{sec:Background}
    
{\it Artificial intelligence} (AI) describes both a property of computer systems---i.e., displaying intelligent behaviour\footnote{Where `intelligence' is understood as `an agent's ability to achieve goals in a wide range of environments' \citep{Legg-Hutter-2007}.}---as well as a set of techniques that researchers use to achieve this property \citep{Gabriel-2020}. Machine learning (ML) is a branch of artificial intelligence wherein algorithms use data to learn gradually. The three main approaches to machine learning are supervised learning, unsupervised learning, and reinforcement learning.

In supervised learning, the algorithm is trained on labelled examples; this is the {\it training data}. The model is evaluated on how well it generalises what it learns from the training data to previously unseen examples (called test data). For example, one might train an image-recognition model on millions of labelled images; if it performs well, it should correctly label novel images not in the training set. In unsupervised learning, an algorithm learns underlying patterns or correlations from unlabelled data. For example, recommendation systems group users together based on their viewing patterns to recommend similar content. Reinforcement learning depends upon sparse rewards for actions. For example, a chess game has a reward of $+1$ if a player wins, $-1$ if a player loses, and $0$ if a player draws. A reinforcement learning model can learn to play chess merely by playing hundreds of thousands of games and receiving a reward at the end of each game.

Machine learning techniques stand in contrast to {\it symbolic systems} (or `good old-fashioned AI'), where explicit, hand-crafted rules had to be hard-coded into the machine---typically in the form of complex, nested if-then statements. The current driver of AI research is so-called {\it deep learning}, an approach to machine learning that utilises deep neural networks modelled (roughly) after neurons in the human brain. Deep learning uses layers of algorithms to process data---information is passed through each subsequent layer in a neural network, with the previous layer's output providing input for the subsequent layer. One of the key advantages of deep learning techniques is that they do not require the heavily hand-crafted features used by traditional machine learning methods. Deep convolutional neural networks are differentiated from their shallow counterparts primarily in terms of their depth, heterogeneity, and sparse connectivity \citep{Buckner-2019}.
    
In the early days of machine learning, researchers could focus on the fundamental aspects of their work without much concern for the social or ethical consequences of these systems because they were relatively encapsulated in the confines of the research lab. However, \citet{Luccioni-Bengio-2019} highlight that these algorithms are increasingly being deployed in society due in part to the promise of the unprecedented economic impacts of ML applications \citep{Bughin-et-al-2018, Szczepanski-2019, Russell-2019}. Consequentially, these systems' ethical and social impacts have started to be examined by researchers from a wide range of disciplines.

Since 2016, emerging technologies' ethical and social consequences have just begun to have a more prominent role in AI research more generally \citep{Christian-2020}. One worry is that algorithms may latch on to spurious correlations (overfitting the training data), which can lead to behaviour that we would call, e.g., sexist or racist (even if implicitly so) when performed by a human. Another slightly different worry is that the data being used to train an algorithm is inherently biased.\footnote{See discussion in \citet{Green-2019, Johnson-2021}.} Hence an optimally-trained model will perform exactly how it should, given the data it has received, again leading to potentially problematic patterns of `behaviour'. A final worry is that the data is simply incomplete or unrepresentative, meaning that the trained system may be `oblivious' to certain (potentially huge) swaths of society. 

For example, a health algorithm that used health {\it costs} as a proxy for health {\it needs} learned to be biased against Black patients in the United States \citep{Obermeyer-2019}. Similarly, a recidivism-prediction algorithm is twice as likely to have false positives (labelling individuals as high risk for re-offence when they are in fact low risk) for Black individuals and twice as likely to have false negatives (labelling individuals as low risk for re-offence when they are in fact high risk) for white individuals \citep{Angwin-et-al-2016}. 

To put it another way: at {\it best}, a sophisticated ML system can be an efficient tool that reflects exactly what the data tell it about the real world, meaning that if the data are biased---which they typically are---then the ML system will perpetuate the existing biases or inequalities in the social system that gave rise to those data. 

%As mentioned in the introduction, cultural code-switching is particularly common among minority communities. In the next section, we describe cultural code-switching and provide some examples. Later in the paper, we will discuss how the biases of AI technologies---especially in the field of natural language processing---come to bear on cultural code-switching. 

%\section{Natural Language Processing}
%    \label{sec:AI-NLP}

    %As we have seen, there are both linguistic and non-linguistic aspects inherent to cultural code-switching phenomena. 
    
    As mentioned in Section~\ref{sec:Introduction}, cultural code-switching is especially common among minority communities, implying differential values and risks for individuals in minority groups compared with those in majority groups. Thus, biases in AI technologies can come to bear in significant ways on cultural code-switching (which we discuss in more detail in the next section). Owing to the communicative nature of  code-switching, the field of artificial intelligence in which an analysis of code-switching will be most relevant is {\it natural language processing}, which studies natural language interactions between computers and humans. 
    
    Some of the key areas of study in NLP research include speech recognition, which may involve recognising spoken language and translating it into text format; natural language understanding or interpretation, which is necessary for, e.g., question-and-answer interactions or machine translation; and natural language generation, which outputs text. NLP has many important applications, including virtual assistants, spam detection, speech recognition, named entity recognition, sentiment analysis, question answering, automatic text summary, autocomplete, predictive typing, relationship extraction, and machine translation, to name a few.
    
    For many years, researchers have utilised shallow models (such as support vector machines or logistic regression) or explicit symbolic representations (such as first-order logic) to solve particular NLP tasks. However, the advent of {\it deep learning} techniques has allowed for significant progress in NLP research in the last decade. You are probably familiar with NLP, even if you have not heard of it---NLP models are necessary for consumer-facing applications like Google Assistant, Siri, or Alexa to work. 

    Consider the last of these. Amazon's Alexa is a virtual assistant AI system that makes significant use of NLP models. It constantly monitors signals in the environment and processes them to minimise ambient noise (such as the conversation on the television) that is not relevant to its task. This pre-processing is always going on in the background. For the software to turn on, it requires a `wake word'---e.g., {\it hey Alexa!} Thus, before Alexa even does anything, it must be capable of differentiating speech from background noise, detecting very specific signals, and responding to them. Once Alexa wakes, it converts the recorded audio to text by analysing certain features like frequency and pitch---this is {\it speech recognition}. Once the speech has been converted to text, Alexa must interpret the meaning of (at least parts of) the text to respond appropriately---this is natural language {\it understanding}. Once the request is processed and understood, Alexa may need to produce a `voiced' response. 

    However, NLP research generally starts with the assumption that language is solely about the transfer of information---i.e., the {\it content} of the message. As mentioned in the introduction, there are many extra-linguistic factors involved in cultural code-switching, so this emphasis on content is myopic. Even granting the view that language is primarily a way of communicating or transferring information, language is but one way of doing so. 

Two of the leading language models in machine learning are {\it Bidirectional Encoder Representations from Transformers}---also known as BERT \citep{devlin2019bert}---and {\it Generative Pre-trained Transformer 3}---also known as GPT-3 \citep{brown2020language}---which is the third iteration of an autoregressive language model, produced by OpenAI. Let's consider the latter of these.

The full version of GPT-3 makes use of around 175 billion parameters---two orders of magnitude larger than its predecessor, GPT-2, and one order of magnitude larger than the next-largest NLP model, Microsoft's {\it Turing NLG}.\footnote{Since the release of GPT-3, this model has been surpassed in size by Google's 1.7 trillion parameter language model. This show of one-upmanship was criticised by \citet{bender2021dangers}, who highlight, among other things, the environmental costs of training large language models.} The majority of training data for this model (around 60\%) comes from a dataset called {\it Common Crawl}---a corpus generated from crawling the Internet---which, as of April 2021, contains around 320 {\it tebibytes} (TiB) of data taken from 3.1 billion web pages. GPT-3 used a filtered version of the Common Crawl dataset, resulting in approximately 14 billion tokens (though they do not provide details of how the Common Crawl dataset was curated). 

To give a sense of this dataset's scale, GPT-3 was also trained, in part, on the {\it entire} English-language Wikipedia, which consists of 3 billion tokens, and accounted for only around 3\% of the training data for the model. Additional datasets used to train GPT-3 include WebText2---consisting of the text of web pages from all outbound Reddit links from posts with three or more `upvotes'---and Books1 and Books2---which are two internet-based books corpora. 

Note, however, that the Internet---and therefore, the data used to train GPT-3---is predominantly {\it English}: around 45\% of HTML pages in the Common Crawl dataset have English as their primary language. It is estimated that English is used by around 62\% of all the websites whose content language we know, with the next most predominant language being Russian, at 8\% \citep{W3TechS}.

Thus, there is likely an anglocentric bias in the data used to train large language models. However, it is perhaps relevant to note that it is very difficult to make claims about the training of these models with any degree of certainty. For example, OpenAI does not describe the filtering process for the Common Crawl dataset, and the WebText2 dataset and the books corpora are not publicly available and lack official documentation \citep{bandy2021addressing}. 

Furthermore, as others have pointed out, the Common Crawl dataset will overrepresent those populations with higher internet usage rates \citep{luccioni2021whats}. And, those countries with the highest internet penetration rates are European (Northern, Western, and Southern) and north-American, with the lowest rates coming from Africa and South Asia \citep{Statista-2021-Internet-Penetration}. This discrepancy leads to a `digital language divide' on the Internet---despite there being around $7000$ languages spoken worldwide, only a few hundred of these are represented on the Internet \citep{Young-nd}. Thus the datasets being used to train these language models are inherently biased, and there has (to date) been relatively little analysis of these biases.

In some ways, the simple case is the one where an algorithm is trained on a specific language; yet, the technologies are highly biased in favour of anglophones. More difficult cases involve particular accents or dialects, and there is already overwhelming evidence that NLP systems are biased---this time in favour of a particular {\it type} of Anglophone: one whose idiolect resembles something like `broadcast English' \citep{Harwell-2018}. Thus, the ubiquity of code-switching highlights a further difficulty that these already inherently flawed systems will face. 

The biases that we have been discussing in NLP models have been primarily linguistic in nature. However, as mentioned in the introduction, there is a broader sense of code-switching that we take to be pertinent here: namely, cultural code-switching. In the next section, we discuss how the pressures to code-switch---to adapt, minimise, suppress, mask, hide, or conform---are especially pressing among already disadvantaged populations. Often, code-switching is a skill that is required to gain access to safety and opportunities for improving one's material conditions. But, at what cost?

 %Furthermore, internet communities supported by anonymity and particular norms can amplify toxic discourse that would not be found in mainstream corpora (Massanari, 2017) often exacerbated by the well-documented `online disinhibition' phenomenon where users find themselves more likely to engage in anti-social behaviours due to the lack of immediate social feedback (Wachs et al., 2019; Mathew et al., 2019; de Lima et al., 2021). This can further perpetuate the lack of diverse, representative language models that can adequately mirror society beyond the boundaries of internet communities.

%{\color {red} Note about connections between language and culture} \cite{Nowak-2020}

%{\color {red} Transition from simple to complex cases}

%% FIX TRANSITION

\section{Cultural Code-Switching}
\label{sec:Cultural-Code-Switching}

    Cultural code-switching is ubiquitous in human interactions: it is used to switch between formal and informal contexts; to wield power and to signal asymmetric authority relations \citep{Popa-Wyatt-Wyatt2018, Tirrell2012}; and can express a sense of agency over one's social identity, and serve to build and reinforce a sense of community and solidarity by signalling in-group affiliation \citep{Dembroff-Saint-Croix-2019}. Since cultural code-switching takes account of the socio-pragmatic aspects of communication, it includes extra-linguistic cultural artefacts (in addition to linguistic ones). For example, consider how a student may modulate their tone of voice or change the register of their speech by using honorifics like `ma'am', `sir', `doctor', or `professor' when interacting with faculty. Thus, instead of the narrow sense of altering some linguistic features of communication, we take code-switching more broadly to include non-linguistic communicative signals which convey information---e.g., about oneself and one's relation to others in a given social context.

    Why do we code-switch? The reasons for code-switching are vast and complex. One important use of code-switching is to signal in-group membership and affiliation. Consider how teenagers (in a broadly North American context) will often use various slang terms and expressions when talking with each other. A quick search through Urban Dictionary, a crowd-sourced online lexicon used to archive slang words and phrases, reveals a host of in-group expressions of this sort---e.g.,  `lit', `salty', `bae', `simp', `noob', `ghosting', `dank', `taking an L', and much more. If you don't know what these expressions mean, that is part of the point. When young people use these expressions among friends, they are code-switching (whether they realise it or not). They are signalling to one's audience that one is `in the know', `one of them', or as they might colloquially put it `cool'.

    Consider another example from the LGBTQ2+ community, particularly the sub-culture of drag queens \citep{Mattel-2018}. If you go to a drag show, there are a range of unspoken cultural norms and expectations that are operative. If it is your first time, you might be taken aback by the playful, mock-impolite, bantering style that drag performers often use. In some cases drag queens might be `throwing shade': a kind of indirect insult toward other drag performers or audience members. This phenomenon is discussed in Jennie Livingston's \textit{Paris Is Burning} (1991), a documentary on drag culture in New York City. Dorian Corey, a drag queen interviewed in the film, gives the following example of throwing shade. %
    \begin{quote}
        If I were to say in a terribly condescending voice, `Oh honey, I'm so glad you saved up to buy those glasses', that's blatant shade. I didn't insult the glasses, or you, directly. It's implied by my voice and the context of what I said. You know they're ugly.
    \end{quote}%
    It is important to note the central role that tone of voice and context play in this example. This is a key issue which we will return to in discussing the relationship between the phenomena of cultural code-switching and human-to-AI interactions. More broadly, this style of communicative engagement has been identified as a kind of pro-social in-group communication, which builds community and solidarity within LGBTQ2+ communities. Additionally, mock impoliteness is also a rhetorical strategy that can be deployed as a form of self-protection; it is a way of preemptively developing a thick skin against bigotry and discrimination \citep{McKinnon-2017, Olivia-et-al-2021}.

    As another example, cultural code-switching in the context of race plays a major role in the plot of Boots Riley's black-comedy film, {\it Sorry to Bother You} (2018). The film centres on a young, Black man named Cassius Green (played by Lakeith Stanfield) who gets a job as a telemarketer and finds major success after he is coached by an older (also Black) co-worker (played by Danny Glover) to use his `white voice'. White actors voice the `white voices' for these characters. One of the major themes of this film is how being successful in a predominantly white culture requires conforming to the expectations of the dominant group---in this case, a marginalised Black man `sounding' white to achieve success at work. According to \citet{Toure-2018}, `putting on a white voice' means to embrace `the ease that white privilege brings. It means sounding as if you’re entitled to the good life. It means feeling calm way down in your soul. It means never having to be afraid someone will call the police on you just because you’re breathing.' However, the film also explores the social consequences of code-switching with respect to one's in-group: Green ends up alienating (or being alienated from) his friends, family, and social in-group.

    This resonates with a broader experience of \textit{double alienation} that individuals from marginalised groups often confront as they find themselves existing between disparate social worlds with distinct and often incompatible cultural norms, values, and expectations. \cite{Morton-2019} describes how first-generation scholars often confront ethical compromises in the process of trying to gain upward social mobility through pursuing a degree in higher education. In conforming to the dominant norms operative in institutes of higher education (especially within elite institutions with predominately upper-class and white student and faculty populations), students from first-generation or low socioeconomic backgrounds often feel isolated and less able to connect with family and loved ones. This is perhaps especially true within fields like philosophy, where first-generation and low-income academics may be unable to fully share their academic life with family and loved ones, who may not understand the point or value in pursuing a career as an academic philosopher. In a recent paper, Morton, who is a first-generation scholar, says of her family that: `We love each other, but I am now part of a world whose logic is mysterious to them' \citeyearpar[10]{Morton-2021}. Thus, the process of code-switching can give rise to serious ethical trade-offs and compromises.\footnote{For further discussions of related issues see \cite{Morton-2019}, and for discussion on the case of academic philosophy specifically, see the recent collection of papers in \cite{Falbo-Stewart-2021}.}

    More generally, one key use of code-switching is cultivating a sense of belonging and community with those in one's social group. One code-switches to express a shared social identity with others; to make others feel comfortable, at ease, or `at home' in a given social environment. This is just one---distinctively \textit{positive} and \textit{self-affirming}---use of code-switching. Other uses of code-switching are starkly different.

    In some cases, one might code-switch not as a means to increase a sense of community with others but rather to navigate an unfamiliar and potentially unwelcoming (or even hostile) social environment. In this sense, the `masking' behaviours that autistic individuals may perform---faking eye contact, mirroring gestures and expressions, scripting conversations, disguising `stimming' behaviour, enduring sensory discomfort, etc.---are a type of cultural code-switching \citep{Hull-et-al-2017}, which may be employed in a range of social situations---including feeling safe in a culture that has, historically, highly stigmatised autistic individuals \citep{Silberman-2015, Donvan-Zucker-2016}.  
    \par Importantly, for our purposes, each of the behaviours just described is {\it non}-linguistic. In this case, regular practice of cultural code-switching can have inherently negative side effects, including increased stress, anxiety, depression, and exhaustion or `autistic burnout' \citep{Bargiela-et-al-2016, Cage-et-al-2018, Cage-Troxell-Whitman-2019, Raymaker-et-al-2020}, or loss of identity and increased risk of suicidal thoughts \citep{Cassidy-et-al-2020}. In this case, too, certain intersectional groups may feel differential social pressure to code-switch---several studies have suggested that people who identify as women are more prone to code-switching than those who identify as men, leading women to be misdiagnosed and partially causing the gender gap in autism diagnoses \citep{Gould-Ashton-Smith-2011, Dworzynski-et-al-2012, Lai-et-al-2015, Hull-et-al-2019}.

    Thus, code-switching is a crucially important skill for negotiating the social dynamics of high-stakes environments---e.g., interactions with the police, or in courtrooms, educational institutions, job interviews, and more. Moreover, the pressures to code-switch are not felt uniformly but are disproportionately experienced by members of historically marginalised groups, who often must assimilate to dominant cultural norms to gain social acceptance and access  certain resources or benefits or avoid potential harms and mistreatment.

    For example, empirical studies and first-person testimonies have shown that Black women with natural hairstyles (e.g., Afros, braids, twists) are less likely to get job interviews than Black women with straightened hair, and especially compared to white women \citep{Koval-et-al-2020}. Black women with natural hair received lower scores on assessments of `professionalism' and `competence', and they were less likely to be invited for job interviews as a result. Most notably, this occurred when the norms of the job required a more conservative and formal dress code. One striking example is the 2017 case of Destiny Tompkins, a young Black woman who was told by her manager at Banana Republic (a white woman) that her braided hair was `unkempt' and `too urban' for the store's image. The manager told her to take out her braids or else she would stop scheduling her for shifts \citep{Samotin-2017}. It is reasonable that Black women might decide to code-switch---e.g., by straightening their hair for job interviews---in order assimilate to norms of `professionalism' in white-dominated spaces. This is often needed to ensure job security and social acceptance.
    
    Of course, code-switching in high-stakes settings, especially where particular opportunities for social mobility and benefits reside, involves more than changes in one's physical appearance. \cite{Dembroff-Saint-Croix-2019}  emphasise the relationship between cultural code-switching and what they call \textit{agential identities}. They define agential identities as `the self-identities we make available to others---they bridge what we take ourselves to be with what others take us to be' \citep[572]{Dembroff-Saint-Croix-2019}. They discuss how code-switching is an effective way to negotiate and switch between \textit{entire social identities}, in some cases moving between genuine identities and merely apparent or superficial social identities. 
    
    %
    %Conforming to dominant cultural norms may require one to change how they talk, e.g., by downplaying a Southern accent, which may be taken as indicating a lack of education in the specific context \citep{Kinzler-2013-et-al}; or what one wears, e.g., consider how a Muslim woman might removing her hijab for a job interview. Also, consider how women in academia, especially within fields dominated by white cis men (e.g., STEM or Philosophy), may deliberately choose not to wear overly feminine clothing and limit the amount of makeup or fashion accessories they wear at conferences or job interviews. Some women do this out of fear that they will be considered `frivolous' or `ditsy', and hence, less of a `serious intellectual' or `professional scholar' \citep{Tsaousi-2020}. 
    
    At the most general level, cultural code-switching influences one's overall mode of existence in social space---how one chooses (or is forced) to exist in the social world. Particular social environments and cultural contexts that incentivise cultural code-switching may be sites of epistemic oppression where silencing thrives.

\section{Cultural Smothering and Double Binds}
    \label{sec:Cultural-Smothering}

    When one is forced to code-switch in the ways that we discussed in the previous section---to mask, blend in, or otherwise express a superficial identity or persona to be accepted, make a living, or avoid harms---it results in a form of self-censorship or self-silencing. This phenomenon is akin to what \cite{Dotson2011} has analysed as \textit{testimonial smothering}. On Dotson's analysis, testimonial smothering is a form of self-silencing that occurs when a speaker truncates or self-censors their testimony because their audience is taken to lack the required competence needed to properly understand what one is saying. This amounts to a kind of \textit{epistemic violence} when the testimonial recipient exhibits \textit{pernicious ignorance}. Pernicious ignorance, according to Dotson, is a form of reliable ignorance, or reliable insensitivity to the truth, that results in a harm.\footnote{Also see, for example, \cite{Dotson2014, Medina2013, Pohlhaus2012} and \cite{Fricker2007} for related discussion.}

    Dotson's analysis of testimonial smothering highlights the importance of testimonial competence---the ability of an interlocutor to \textit{get what you mean} (in the way you mean it)---for successful communicative exchanges. Furthermore, this analysis describes how particular speakers, especially those who are testifying from disadvantaged positions, are prone to vulnerabilities and harms when their audience lacks such a competence. \citet{Dotson2011} writes
        \begin{quote}
            Speakers are vulnerable in linguistic exchanges because an audience may or may not meet the linguistic needs of a given speaker in a given exchange. \ldots [T]o communicate we all need an audience willing and capable of hearing us. (238)
        \end{quote} %
    Moreover, engaging in communicative exchanges where one's audience lacks testimonial competence (or where it is unclear if one's interlocutor is competent) can be very risky and potentially unsafe \citep[244-245]{Dotson2011}. For example, drawing upon the work of \cite{Crenshaw-1991}, Dotson explains how Black women's testimony about sexual violence perpetrated by Black men is potentially unsafe or risky in social contexts where it may be interpreted as reinforcing racist stereotypes. 

    Dotson’s analysis focuses on \textit{testimonial} exchanges---how testifiers from oppressed groups are often forced to capitulate to the testimonial incompetency of their audiences by altering their testimony to include only that which is likely to be given proper uptake. However, with cultural code-switching in view, we can broaden Dotson's analysis to apply not only to the truncating or smothering of testimony but also the smothering of aspects of one's cultural identity more broadly. This is salient in cases of forced code-switching, where one masks aspects of their culture in response to dominant norms and pressures within the social environment, especially where code-switching is required to gain some benefit or to avoid harms. In such cases, one may not only smother their testimony, but also engage in a broader form of what we will refer to as \textit{cultural smothering} \citep{Falbo-2021}. When cultural code-switching behaviours manifest as cultural smothering, they take on a form of self-silencing.

 The pressures to engage in cultural smothering through code-switching arise in non-ideal choice situations. Such situations often have the structure of a \textit{double bind}, which is described as a `situation in which options are reduced to a very few and all of them expose one to penalty, censure or deprivation' \citep[2]{Frye-1983}. For example, when deciding whether to have (heterosexual) sex, Frye discusses how women are often susceptible to social criticism and scrutiny no matter what they do. If they have sex, they are prone to be called `easy' or a `slut', but if they do not, they are seen as `prudish', `uptight', or `frigid'. No matter what a woman chooses, she loses. The disadvantage that double binds give rise to is a function of their structural properties; therefore, they are sites of systemic oppression.  They result from, reinforce, and sustain, oppressive systems and unjust social arrangements.

Recently, \cite{Hirji-forth} has argued that what makes double binds unique and effective vehicles of oppression, compared to other difficult choice situations, is that `whatever an agent does necessarily undermines their own objective interests' (3). Hirji develops the notion of an `imperfect choice', which is a kind of choice situation where it's  impossible to advance one's interests and where one will inevitably be worse off as a result. Imperfect choices constrain one's agency while nonetheless leaving aspects of one's autonomy intact. One is still free to choose, but this freedom is hollow and illusory because all available options undermine one's interests. Moreover, in so choosing, one is forced to be complicit in their own oppression. Double binds present a series of options, all of which are non-ideal and further contribute to systems of oppression that function to make one, and members of one's social group, worse off. 

How does this relate to cultural code-switching? In cases where one must code-switch to gain some advantage or to avoid some harm in the social environment, one typically faces an imperfect choice. They are in a double bind. On the one hand, one might choose to code-switch. If so, one engages in a form of self-silencing or self-censorship: they culturally smother in response to (likely or actual) pernicious ignorance. They conceal, truncate, mask, or alter aspects of their social identity, presenting an edited version of themselves that accords with dominant cultural standards. On the other hand, if one decides {\it not} to code-switch, they resist dominant cultural norms and expectations within the social environment. But, by not code-switching, one risks social exclusion, unemployment, forms of material deprivation, and even safety (as was discussed in Section~\ref{sec:Cultural-Code-Switching}). 

No matter the choice, one's interests are undermined. By code-switching, one is forced to be complicit in reinforcing dominant cultural norms, further entrenching broader patterns of inequality. But, by not code-switching one risks losing important material goods and opportunities for social advancement.

%The pressures to code-switch in these circumstances put into sharp focus a tension between reconciling one's own identity and where one comes from---one's family and home life, one’s community, values, and importantly, one’s core sense of self---with where one is aspiring to go.

The pressure to code-switch highlights how members of marginalised groups must navigate unfamiliar and potentially hostile social environments to pursue upward mobility. It is also important to remember that code-switching can be \textit{exhausting}, especially when it manifests as cultural smothering. It involves a great deal of psychologically taxing work in which one's comparatively privileged counterparts simply need not engage.

\section{Tentative Lessons and Concluding Remarks}
\label{sec:Conclusion}

AI systems are increasingly ubiquitous and more complex. So too are the social issues to which they give rise. This includes not only more direct human-to-AI interactions via smartphones, chatbots, or digital voice assistants but also how these technologies are implicated `behind the scenes' through social media platforms, streaming services, predictive search algorithms, video games, apps for rideshare programs and online dating, email communications, banking and finance, e-commerce, and more. These advancements have undoubtedly led to many positive changes, making previously arduous tasks much more streamlined and efficient. But, at the same time, these technologies can hinder social progress.

As is now well known, AIS encode biases. This is unsurprising. These systems are not getting their training data in a vacuum. Instead, they are sourcing it from places like the Internet,  which reflects society's myriad biases and prejudices, and produces selection effects favouring certain privileged populations (e.g., English-speakers, those who have Internet access, etc.) over others. As these technologies become increasingly advanced and indispensable to everyday tasks, researchers must be cognisant of how social structures that maintain unjust social arrangements are reproduced, reflected, or potentially made more potent and harmful within these technologies and their implementation.\footnote{This discussion also raises the question of whether we should really want AI technologies, for instance, voice assistants like Alexa and Siri, to be very human-like. One interesting development is the creation of Q, a genderless voice assistant. You can try it out yourself: \href{www.genderlessvoice.com}{www.genderlessvoice.com}.}

These insights underscore the need to attend to broader structural inequalities and mechanisms of oppression that reach well beyond AI technologies. It suggests that the solution to these issues does not rest within some manipulation or re-configuration of training data or some alternative algorithm. The problems we identify are not fundamentally problems resulting from AI technologies and, as a result, won't give way to technological solutions.  As \citet{O'Neil-2016} has argued, `Big Data processes codify the past. They do not invent the future' (204).  Importantly, more data and more accurate or efficient algorithms will not solve any of the problems that we have described. Even if an algorithm had a perfect model of the world, the structures that disadvantage certain groups over others would be reflected in the data. This is because the problems that arise when training and implementing AI technologies in a non-ideal and unjust social world are due to the very structure of that world. As a result, the issue is not whether the data are incomplete or unreflective of the real world. Even (perhaps especially) if they are complete and perfectly reflective, these problems will persist. Moreover, these technologies feed back into society, creating more biased data as input, thus giving rise to a negative feedback loop, of the type described by \citet{O'Neil-2016}. Accordingly, the solution cannot be merely technological because the problem is not merely technological. A responsible first step is to be aware of structural injustice in the world. The point of intervention is not in changing our data to be better---our data will be better when the world is better---but when inequalities persist, so too will inequalities in data emerge to reflect them. All technological progress does is instantiate these inequalities more efficiently.

We have highlighted one important way these technologies can potentially be sites for epistemic oppression: by imposing imperfect choice situations or double binds onto non-dominant social groups. Just as in the case of human-to-human interactions, where patterns of silencing emerge in response to the pernicious ignorance of one's audience, so too can pernicious ignorance be encoded into AI technologies, further deepening patterns of silencing on a potentially grander scale and at a more rapid pace. If technologies involving NLP and human-to-AI-interactions require speakers to conform to dominant expectations and norms---or, in other words, if such technologies encode pernicious ignorance---then non-dominant groups will need to engage in cultural smothering in order to interact with these technologies.

\iffalse
    \newpage
    \section{NOTES (Travis)}
    
    \begin{itemize}
        \item Say something about default-white (male, {\it anglo}) assumptions in ML research; the last of these is particularly important for present purposes.
        \item Emphasis in NLP is on syntax and semantics (cite examples or survey); pragmatics are difficult but important for the context-sensitive nature of language which we discuss here.
        \item Mention NLP programmes that flag drag queen speech (and the speech of other queer communities) as `toxic' (i.e., insensitive to context).
        \item Cite Gabbrielle Johnson's paper on value-free machine learning.
        \item Cite DeepMind paper on `fairness for unobserved characteristics'. 
        \item PAPER STRUCTURE:
        \begin{itemize}
            \item Introduction - essentially the extended abstract
            \item Discussion of Cultural Code-Switching
            \item Background on SOTA NLP research
            \item Importance of code switching in the context of AI ethics research
            \item Value-ladenness of science
            \item Conclusion
        \end{itemize}
    \end{itemize}
\fi 

\newpage
\singlespacing
\bibliographystyle{apalikelike}
\bibliography{Biblio}

\end{document}